# Survey on Stabilization of Nonlinear Systems via state/output feedback control

"Deguale Demelash abiye"

Email; demelashabiye@csu.edu.cn

"*School of Automation*"

Central South University, Changsha 410083, China

**Abstract:** This survey paper deals with the stabilization of nonlinear systems by analyzing the controlling method in terms of state feedback and output feedback. A brief overview of some literature on how the feedback controller of some dynamic systems in real applications like robots can be applicable is introduced. The aim is to give a direction about the methods of how to solve the state and output feedback stabilization problems for nonlinear systems based on feedback design methods and other theorems like the Lyapunov stability theorem. The feedback controllers can be constructed to ensure the origin of the nonlinear system is whether asymptotically stable or not. The paper is purposely focused on directing the theoretical results regarding the presence of such a feedback controller in stabilization.

## 1. Introduction

In this paper, a nonlinear system can be described as a set of nonlinear equations used to investigate a process or any physical device which cannot be clearly described by a set of any type of linear equations [2]. When engineers design and analyze nonlinear systems in different engineering disciplines like mechanical systems, electrical circuits, control systems, they have to use an extensive range of nonlinear analysis methods. In spite of the fact that these methods have advanced rapidly since the mid-1990s, nonlinear control is still mainly a tough task [1]. As it is known, stabilization of a nonlinear system with output feedback is a problem that needs much consideration and challenges. The traditional output feedback design method is not applicable because of the lack of consistent non-observable linearization non-observability of nonlinear systems. Ever since the output feedback stabilization of planar systems was established and improved [3].

    The design and analysis of reduced-dimensional observers for a nonlinear system have made extraordinary progress in recent years. The problem of stabilization for nonlinear dynamic systems has attracted an excessive deal of devotion from the control community [4]. For stabilization of nonlinear systems under state feedback control is studied in the paper [5]. To achieve stability, dynamic quantization approaches are presented in general. Feedback stabilization of nonlinear systems has been demonstrated

to rely on aspects of closed-loop systems without state quantization, such as input-to-state stability [5].

The main advantage of the dynamic controller in nonlinear systems over other predictor controllers or successive controllers is the existence of simple formulas for the approximation of the asymptotic improvement of the measurement error for certain classes of systems. The numerical predictor, on the other hand, is the predictor for which the influence of measurement mistakes is the most difficult to quantify [6]. The dynamic controller, on the other hand, has the disadvantages of being difficult to implement (one must approximate numerically the solution of the IDEs or equivalent distributed delay differential equations) and the fact that it works only for certain categories of nonlinear systems, like systems with a compact absorbing set and globally Lipschitz systems [6,7].

This survey paper is organized as follows. It contains the related idea of stability and stabilization. Then feedback control systems (both state feedback and output feedback) are introduced in terms of stabilization. Finally, a brief review of how these control systems can be applied in real-world applications is introduced.

## 2. Stability of Autonomous nonlinear systems

The goal of stabilization is to keep the system close to an equilibrium point b. The goal is to create stabilizing control system such that under such control laws, b is becoming an asymptotically stable equilibrium position of the system [25]. Some basic principles and features of nonlinear control systems are described in the reference [24]. Furthermore, the core Lyapunov's stability theory, as well as the ideas of stability and robust stability, are summarized. The given equation bellow describes the state space description of a nonlinear system [24];

$$\dot{x}(t) = f(x(t), t), \ x(t_o) = x_o, \ t \geq t_o \qquad (1)$$

where, $t \geq t_o$ is the initial time, $x : (t_o, \infty) \to R^n$ and $f: D \to R^n$ is a locally Lipschitz map [1] from the domain $D \subseteq R^n$ to $R^n$. Assume that the system in equation (3-1) has an equilibrium point $z \in D$, such that, $f(z) = 0$. So, this can occur if the equilibrium point $z$ is stable.

One of the pillars of automated control and stability for ordinary differential equations in particular is the Lyapunov stability theory. Lyapunov's original theory, which originates from 1892, is concerned with the performance of differential equation solutions under various beginning circumstances. The study of liberations in astronomy was one of the applications that was considered at the time. The focus is on ordinary stability (i.e., stable but not asymptotically stable), which we may describe as robustness with regard to starting circumstances, with asymptotic stability considered only as a corollary [26]. So, to check the stability of autonomous nonlinear systems easily, we can use the suitable Lyapunov function [8]. We can consider a nonlinear autonomous system as follows;

$$\dot{x}_1 = -x_1 + x_2^3$$
$$\dot{x}_2 = x_1^3 - x_2 \qquad (2)$$

To examine the stability of the system by the Lyapunov function, use the suitable Lyapunov function $V = x_1^2 + x_2^2$. The derivative of V is;

$$\frac{d(V)}{dt} = 2(x_1^2 + x_2^2)(x_1 x_2 - 1) \qquad (3)$$

From this, we can conclude that the system is stable if and only if $x_1 x_2 > 1$.

The stability ideas for the generic non-autonomous system defined by equation (3-1) are investigated in [24], with autonomous systems being viewed as a particular instance of equation (3-1).

## 3. Stabilization by feedback control

Process safety plays an important part in the design of cyber-physical and networked control systems. A critically stable control system is required in autonomous vehicles, chemical reactors and robots, and so on. As a result, the feedback controller is built to match required performance with safe stability margins and state restrictions to avoid dangerous states, as well as input constraints that are quite severe [8]. In the literature [18], numerous control design strategies for nonlinear systems have been developed, including stabilization and controller design with nonlinear constraints. Model-based controllers, Model Predictive Control, and intelligent controllers, for example, are offered for trajectory tracking and stabilization with input-output limitations to achieve optimal performance in [13,18]. In addition, a new event-triggered technique for the stabilization of a general nonlinear system is developed [34], in which the sampling error is regarded as an actuator disturbance. This novel technique introduces a milder solvability condition for the event-triggered stabilization problem than the previous one, which treats sampling error as a sensor disturbance. This novel technique was also used to solve the global robust output regulation problem in an event-triggered scenario. Unmodeled dynamics exist in practically all actual nonlinear systems, and they can certainly reduce closed-loop system performance due to variables such as measurement noises, modeling mistakes, and modeling simplifications. It is also one of the primary causes of system instability. Several various ways to handle such systems with unmodeled dynamics employing backstepping or dynamic surface control have been proposed in the literature [19]. Given previous knowledge of the system zero-error steady-state condition and a correct internal model, a systematic stabilization strategy is proposed in [35] for systems whose regulatory error dynamics are subject to rational nonlinearities. The error dynamics are expressed in differential algebra in order to address the synthesis of controller settings via a numerical optimization problem constrained by bilinear matrix inequality constraints.

However, all of these studies are limited to SISO nonlinear systems, in which nonlinear uncertainties are solely state-dependent, limiting their usefulness in real-world systems Predictor feedback is devised and stabilization is proven for the class of nonlinear delay systems with a compact absorbing set in [6], and the concept is later expanded to nonlinear delay systems that can be changed to systems with a compact absorbing set using preliminary predictor feedback. Reference [20] provides a robust model predictive control approach for restricted discrete-time nonlinear systems with unmodeled dynamics, based on a partially known nominal model, to decrease the system's instability and ensure stability. Ref [28] considers a global robust asymptotic stabilization issue (GRS) for cascaded systems with dynamic uncertainties that are not always input-to-state stable (ISS). A recursive Lyapunov design approach is built specifically through induction on the system relative degree, resulting in a smoothly globally stabilizing controller. It can be applied to nonlinear cascaded systems with several different ISS dynamic uncertainty. The suggested design is constructive and results in an ISS-Lyapunov characterization of the closed-loop system in superposition form. There are many applicable results on feedback stabilization (see [22,23]) that have been obtained in recent times. From these results the following two are the most important; The first assures the existence of piecewise analytic stabilizing feedback laws under relatively modest controllability assumptions, whereas the second sets extremely severe required conditions for the growth of stabilizing $C^1$-feedback laws.

One of the purposes of feedback control systems is to stabilize potentially unstable systems. Although some results show that a system is reachable, However, an unstable system's state may be controlled by selecting the right control input, these results were achieved under the following conditions [9]:

- ✓ Unrestricted control is required.
- ✓ The system has to be accurately defined (i.e., it must have an accurate model of it)
- ✓ The initial state needs to be known precisely.

The problem with unstable systems is that they are difficult when assumptions like the ones above fail to hold true. Even if the first assumption is considered to be true, modeling faults such as incorrectly represented dynamics or incompletely modeled disturbances will surely exist (so, it will be violating the second assumption). Again, if we assume that the dynamics are correctly modeled, the system's starting state is unlikely to be precisely understood (will be violating the third assumption). As a result, it is evident that we require continuous input on the status of the system in order to have any hope of stabilizing an unstable system. Feedback may increase the performance of a stable system (or, if it is poorly chosen, it can lower the performance and potentially create instability.

The important question to do this may be "How can we then create feedback controls that stabilize a system?" To answer this, we must first consider the types of feedback variables available to our controller. In general, there are two sorts of feedback; These are state feedback and output feedback.

## 3.1. State feedback control

The design of feedback controllers such that particular outputs of a given system in predefined reference trajectories is a basic topic in control theory. In an actual case, this control aim must be met despite a slew of factors that might lead the system to react in ways that aren't predicted. These occurrences might be endogenous, such as parameter changes, or external, such as new unwanted inputs influencing the system's behavior [14]. A popular approach to controlling a non-linear system is to linearize the system around the desired value and then construct the controller using linear feedback control methods [10]. In the paper [10, 11], For single-input non-linear systems, a design strategy for state-feedback controllers is presented. The approach makes use of the nonlinear system's conversions into 'controllable-like' canonical forms. The eigenvalues of the linearized closed-loop model are invariant with regard to any fixed operating point in the non-linear state-feedback that results. A comparison of alternative non-linear system transformation strategies is also offered. To demonstrate the design method, an example and simulation results of several control schemes are presented in [11]. The goal is to create controllers that can guide a nonlinear system from its initial state towards a specific area of state space. It is demonstrated that state feedback controllers may be created by solving a series of inequalities derived directly from quantities in the equation of the controlled system. The findings are used to create state feedback controllers that produce robust asymptotic stabilization for a broad range of nonlinear systems [12].

More recently, a low-pass filter in the control loop has been proposed as a safe implementation of delay-distributed control principles. In [13], the state feedback for nonlinear systems via constant input delay is described. The goal of [13] this research was to propose a state feedback rule that would allow the system to be stabilized in a closed loop and, lastly, to expand the finite spectrum assignment technique that is already available for linear input delayed systems. Additionally, the proof that all states of the time delay cascade system exist is done in Ref [27] by creating suitable Lyapunov-Krasovskii functionals to the origin while retaining the closed-loop system's boundedness. The subject of global stability is investigated using partial state feedback for a type of time-delayed cascade system. A delay-free, dynamic partial state feedback compensator is aimed to accomplish global state regulation under acceptable ISS circumstances on inverse dynamics [27]. A basic framework is presented in the paper [36] that systematically turns the robust output regulation problem for a general nonlinear system into the robust stabilization problem for an adequately augmented system. This generic framework, on the one hand, relaxes the polynomial assumption, while on the other hand, provides more freedom to incorporate recent innovative stabilization techniques, so laying the groundwork for systematically addressing robust output control with global stability.

In [28], For nonlinear cascaded systems with input-to-state stable (ISS) dynamic uncertainty, a global robust asymptotic stabilization (GRS) issue is investigated. The presented conclusions apply to systems with various GRS and ISS combined dynamic uncertainties, as well as to systems with partially limited and partially unbounded acceptable growth rates. In comparison to the single-type ISS, this situation is far more intricate and difficult. Ref [29] constructs a family of nonlinear controllers that handle the robust input-to-output stabilization issue with a defined gain using a recursive method. When two unique subsystems are joined, the RISS trait is kept under minimal gain situations, according to the report.

### 3.2. Output feedback control.

All control approaches that simply use the measured output(s) for input signal generation are referred to as output feedback. There are no intermediate signals detected, such as nonoutput states or external signals [15]. Output feedback stabilization of nonlinear systems, as is well known, is an issue that gets a lot of interest and has a lot of challenges. The typical output feedback design method, as described in [12], is not suitable due to the lack of consistent non-observability and non-observable linearization of nonlinear systems. Since the output feedback stabilization of nonlinear systems was established and proposed, and a global finite-time stabilization by output feedback for a class of nonlinear systems was studied in [3,16], a global finite-time stabilization by output feedback for a class of nonlinear systems has been proposed. Ref [39] studied the consensus problem of a class of multi-agent systems (MASs) with linear dynamics, where each agent can use the given measurement output relative to its neighbors at certain sampling instants. The network is considered to have switched but jointly-connected topologies, which means that the union of the associated graphs over a given time interval always contains a spanning tree. Under this assumption, [39] proposes a linear sampled measurement output feedback controller to attain full state consensus for a suitably short sampling interval.

Ref [31] is the first to investigate the challenge of global stabilization through output feedback for a type of uncertain nonlinear systems with unknown growth rates. Because various design parameters in the controller are frequently tied to unknown growth rates, designing an output feedback controller for these nonlinear systems is normally difficult. A global adaptive output feedback controller may be explicitly created using universal control concepts, allowing all states of unstable nonlinear systems to be controlled to zero. Because some nonlinear systems have the unboundedness observability quality, it is recognized that sufficient growth conditions should be placed on the nonlinearities to globally stabilize nonlinear systems through output feedback. In most circumstances, the nonlinearities must only rise linearly in relation to the unmeasurable states. The semi-global robust output regulation problem for nonlinear affine systems in normal form is studied in [40]. A similar topic was previously investigated under the assumption that the solution to the regulator equations is polynomial. We relaxed the polynomial assumption on the solution of the regulator equations by applying the nonlinear internal model approach. [41] also addresses the output regulation problem of singular nonlinear systems using normal output feedback control.

In Ref [33], for a type of uncertain nonlinear systems influenced by both linear and higher-order nonlinearities multiplied by an output-dependent growth rate, the challenge of global robust stabilization via output feedback is studied. The current contribution has two components. One approach is to use homogeneity and dominance to manage polynomial growth situations without introducing an observer gain updated law. Another is the creation of a recursive design algorithm for the construction of reduced-order observers, which is not only interesting in and of itself but also has a valid counterpart that is effective when dealing with strongly nonlinear systems, even when uniform observability and smooth stabilization are lacking. In addition, a dual-observer strategy for global output feedback stabilization of nonlinear systems [16, 17] is observed. For all beginning circumstances in a given compact set, Ref [21] tackles the problem of finding an output feedback law that asymptotically leads to zero specified outputs while preserving all state variables constrained. The challenge may be seen as an extension of the traditional problem of semi-globally stabilizing a controlled system's paths to a compact set.

In recent years, much attention has been paid to the robust servomechanism problem also known as the output regulation problem of the class of nonlinear systems in lower triangular form. Either state feedback or output feedback provided the semi-global answer initially. The worldwide solution based on state feedback was recently provided. However, the global solution via output feedback has long been an unresolved issue. Ref [37] defines a set of solvability conditions for the global robust servomechanism issue for this family of nonlinear systems using output feedback. The polynomial assumption imposed on the solution of the regulator equations is a major barrier to the solvability of the nonlinear robust output regulation problems. This assumption is based on the inability to develop an internal model that can account for more complex nonlinearities than polynomials. It was recently shown that a nonlinear internal model can be built under considerably softer assumptions than the polynomial assumption. According to ref [38], this type of internal model can be utilized to address the global robust output regulation problem for a class of nonlinear systems with output feedback.

Important topics like how to build nonlinear adaptive observers and how to achieve global adaptive stabilization via output feedback have been researched in the literature [30]. Notably, the majority of the findings are limited to a subset of uncertain nonlinear systems with parametric output feedback. Far beyond parameterized output feedback observers, they could be used to regulate nonlinear systems with unknown parameters. Furthermore, such an observer would contain the system's uncertainty and so would be impossible to deploy. Global adaptive stabilization of the uncertain system via output feedback is a non-trivial condition due to the inherent difficulties of this kind [30]. By using error output feedback control, Ref [32] revisits the global robust output regulation (GROR) problem of nonlinear output feedback systems with unpredictable ecosystems. The problem was traditionally addressed by using a linear canonical internal model, and as a result, appropriate adaptive stabilization is required for the augmented system to achieve output regulation. In contrast, [32] develops a novel nonlinear internal model technique that successfully changes the GROR problem into a resilient non-adaptive stabilization problem for the augmented system. Ref [32] also presents the notion of looking for new ways to build internal models that serve global output regulation design. In particular, the internal model dynamics include a convergent estimator for unknown system parameters. In this method, we may separate the augmented system stabilization from the integration of any additional adaptive control rule, simplifying the stabilization operation. Because the suggested internal model completely meets the needed stabilization constraints, the non-adaptive stabilization issue may be handled utilizing a ready approach recently established in [28].

## 4. Conclusion

This survey paper addresses an overview study of literature about the Stabilization of Nonlinear Systems using state output feedback control. The paper gives a direction about articles that studied the feedback controls in the real applications in the mathematical world and engineering. The design and analysis of nonlinear feedback systems have recently undergone an exceptionally rich period of progress and maturation, fueled in large portion by the discovery of certain basic conceptual notions and the identification of classes of systems for which systematic decomposition approaches can result in effective and easily computable control laws.

The method of proving stability by means of vector Lyapunov functionals is an extension of the method of Lyapunov functionals described in some of the literature mentioned in this paper and has a long history. However, proving stability by means of small-gain results is a method radically different from all other methods of proving stability. The obtained small-gain results permit the derivation of novel vector Lyapunov characterizations of global stability notions. A special feature of small gain results is the stabilization of nonlinear systems estimates must be proved by means of other methods (e.g., by means of Lyapunov methods or comparison principles or analytical solutions as mentioned in the reviewed literatures). Another important feature of these methods is that they can easily handle both static and dynamic feedback controller.